\documentclass[twocolumn,pra,showpacs,amsmath,amssymb]{revtex4}
\usepackage{amssymb}
\usepackage{graphicx}

\def\d{_{\mbox{\tiny{DET}}}}
\def\i{_{\mbox{\tiny ISO}}}
\def\m{_{\mbox{\tiny{MAX}}}}
\def\l{_{\mbox{\tiny{LOCAL}}}}
\def\b{\mathcal{B}}

\def\th{^{\mbox{\scriptsize th}}}
\def\one{\leavevmode\hbox{\small1\normalsize\kern-.33em1}}

\begin{document}

\title{General properties of Nonsignaling Theories}

\author{Ll. Masanes$^1$, A. Acin$^2$ and N. Gisin$^3$}
\affiliation{
    $^{1}$Department of Mathematics, University of Bristol, BS8 1TW Bristol, UK\\
    $^{2}$ICFO-Institut de Ci\`encies Fot\`oniques, 08034 Barcelona, Spain\\
    $^{3}$GAP-Optique, University of Geneva, 20 Rue de l'\'Ecole de M\'edicine, CH-1211 Geneva 4, Switzerland}
\date{\today}

\begin{abstract}
This article identifies a series of properties common to all
theories that do not allow for superluminal signaling and predict
the violation of Bell inequalities. Intrinsic randomness,
uncertainty due to the incompatibility of two observables,
monogamy of correlations, impossibility of perfect cloning,
privacy of correlations, bounds in the shareability of some
states; all these phenomena are solely a consequence of the
no-signaling principle and nonlocality. In particular, it is shown
that for any distribution, the properties of (i) nonlocal, (ii) no
arbitrarily shareable and (iii) positive secrecy content are
equivalent.
\end{abstract}

\maketitle

\section{Introduction}

There are two experimental facts that, when considered together,
significantly restrict any possible physical theory that aims at
accounting for them. The first one is the constancy of the speed
of light in any reference frame. This implies that no signal
carrying information can propagate faster than light. More
generally, we refer as {\em the no-signaling principle} the
impossibility of sending information arbitrarily fast. The second
fact, is the existence of correlations between space-like
separated events that violate Bell inequalities
\cite{Bell,aspect}. This means that such correlations cannot be
explained by strategies arranged in the past.
Models accounting for such correlations can be constructed by
assuming some signaling between the correlated events. But this
seems to contradict the first experimental fact. This is the
reason why such correlations are called {\em nonlocal}. Despite
this, physical theories exist that predict the violation of Bell
inequalities and are nonsignaling, an example being Quantum
Mechanics (QM).

QM is not the unique theory consistent with the two mentioned
experimental facts. It is well known that there exist nonsignaling
correlations that are more nonlocal than the ones predicted by QM.
Indeed, Popescu and Rohrlich proved that there are nonsignaling
correlations giving a Bell inequality violation larger than the
quantum mechanical prediction \cite{pr}. This suggests the
possible existence of theories, different from QM, that allow for
Bell inequality violation without contradicting the no-signaling
principle. Although there is no experimental reason to reject QM,
it is highly desirable to know the nature of these alternative
theories in order to "study quantum physics form the outside". In
this article, we aim at providing a unified picture for the {\em
static} part [we do not consider {\em dynamics}] of all such
theories, identifying a series of features common to all of them.

Analyzing these common properties can be very useful in gaining a
better understanding of QM. It is often said that the postulates
of QM do not have a clear physical meaning, especially when
compared with the postulates of other theories, like Relativity or
Thermodynamics. The postulates of QM imply no-signaling [if we
assume locality of interactions], and nonlocality. It was proposed
by Popescu and Rohrlich to consider no-signaling and the existence
of nonlocal correlations as proper physical principles. Could
these two principles, together with other {\em independent}
postulates imply QM? What would these other postulates look like?
For such an enterprise, it is very important to learn all the
consequences that follow from these two principles without any
extra assumption.

From an information-theoretical point of view, it is also worth
looking at a framework more general than QM, as illustrated by
several recent works analyzing the use of nonlocal correlations as
an information-theoretical resource \cite{NSC,infres}. This is of
particular interest in the case of secret communication: there,
the security of a protocol relies on some assumptions on the
eavesdropper capabilities. Usually, it is assumed that her
computational power is bounded, or that her action is constrained
by QM laws. It is then desirable to weaken the strength of these
assumptions as much as possible. In this sense, a secret key
distribution was recently proposed in \cite{crypto} and its
security proved solely using the no-signaling principle. In this
article, we extend the connection between nonlocality and secrecy
at the level of an equivalence. Notice that the fact that a
probability distribution contains secrecy does not imply that it
can be distilled into a secret key (see below).

\subsection{Summary and results}

The article is organized as follows: in section 2 nonsignaling
correlations are introduced, local and nonlocal ones are
distinguished. Special emphasis is made on a particular family of
distributions that we call isotropic, which will prove very useful
in later reasonings.

In section 3, different aspects of monogamy in nonlocal
correlations are presented. In particular, the complete
equivalence between locality and infinite shareability is proven
(section 3.1). In section 3.2, through some examples, we survey
the complex structure of the monogamy relations.

In section 4 we prove that, any nonsignaling theory that predicts
the violation of at least one Bell inequality has a No-Cloning
Theorem. Some additional analysis is made for the case of QM.

In section 5 we prove that, nonsignaling correlations contain
secrecy (in the sense of cost) if and only if they are nonlocal.

In section 6 we review the fact that all nonlocal correlations
must have nondeterministic outcomes. And, in section 6.1 we show
that, the more incompatible two observables are, the more
uncertain their outcomes.

Finally, we conclude with some final remarks, exposing some open
question. Some additional material and proofs is contained in the
appendixes.

\section{Definitions and general frame}

Consider $n$-parties ---Alice, Bob, Clare\ldots--- each possessing
a physical system, which can be measured with different
observables. Denote by $x_k$ the observable chosen by party $k$,
and by $a_k$ the corresponding measurement outcome . The joint
probability distribution for the outcomes, conditioned on the
observables chosen by the $n$ parties is
\begin{equation}\label{correlation}
    P(a_1,\ldots, a_n|x_1,\ldots, x_n).
\end{equation}

One can formulate this scenario in an equivalent and slightly more
abstract way. Imagine that each of the $n$ parties has a physical
device with an input and an output. Just after the $k\th$ party
inputs $x_k$, the device outputs $a_k$, and it cannot be used
anymore. Throughout this article, we assume that inputs and
outputs take values from finite, but arbitrarily large, alphabets:
$x_k\in\{0,1,\ldots, X_k-1\}$ and $a_k\in\{0,1,\ldots, A_k-1\}$.
Notice that, without loss of generality, we assume that all
observables belonging to one party have the same number of
outcomes.

It is useful to look at these conditioned probability
distributions (\ref{correlation}) as points in a large dimensional
space. The set of all these points (\ref{correlation}) is a convex
polytope. Unless no other constraints are imposed,
(\ref{correlation}) can be any vector of positive numbers,
satisfying the normalization conditions
\begin{equation}
\sum_{a_1,\ldots a_n} P(a_1,\ldots a_n|x_1,\ldots x_n)=1
\end{equation}
for all input values $x_1,\ldots x_n$.

\subsection{Nonsignaling correlations}

The $n$-partite distribution $P(a_1,\ldots a_n|x_1,\ldots x_n)$ is
nonsignaling, when the marginal distribution for each subset of
parties $\{a_{k_1},\ldots a_{k_m}\}$ only depends on its
corresponding inputs
\begin{equation}\label{totes}
    P(a_{k_1},\ldots a_{k_m}|x_1,\ldots x_n)=
    P(a_{k_1},\ldots a_{k_m}|x_{k_1},\ldots x_{k_m}).
\end{equation}
It turns out that very few of these conditions are linearly
independent. It was proved in \cite{NSC}, that all conditions of
the form (\ref{totes}) can be derived from the following

{\bf Condition:} For each $k\in\{1,\ldots n\}$ the marginal
distribution obtained when tracing out $a_k$ is independent of
$x_k$:
\begin{eqnarray}\label{ns}
    \sum_{a_k} P(a_1,\ldots a_k\ldots a_n
    |x_1,\ldots x_k\ldots x_n) \\
    =\sum_{a_k} P(a_1,\ldots a_k\ldots a_n
    |x_1,\ldots x'_k\ldots x_n), \nonumber
\end{eqnarray}
for all values of $a_1,\ldots a_{k-1},a_{k+1}\ldots a_m$ and
$x_1,\ldots x_k, x'_k, x_{k+1}\ldots x_n$.

These linear constraints characterize an affine set. The
intersection of this set with the polytope of distributions
(\ref{correlation}) gives another convex polytope. Throughout this
article, whenever we refer to distributions, correlations, states
or points, we always assume they belong to the nonsignaling
polytope.

\subsection{Local correlations}

Local correlations are the ones that can be generated if the
parties share classical information, or equivalently, the ones
that can be written as
\begin{eqnarray}\label{local}
    P(a_1,\ldots a_n|x_1,\ldots x_n) \\
    = \sum_e P(e) P(a_1|x_1,e)\cdots P(a_n|x_n,e). \nonumber
\end{eqnarray}
This subset of correlations is a convex polytope delimited by two
kinds of facets. The first kind warrants that all the components
of (\ref{local}) are positive, and thus, it is not interesting.
Actually, they are already facets of the nonsignaling (and also of
the more general) polytope. The second kind are the Bell
inequalities, which can be violated by nonlocal correlations.
Throughout this article we assume that all Bell inequalities have
been normalized [with a transformation of the form $\b \rightarrow
\alpha\b+\beta$, where $\alpha$ and $\beta$ are real numbers ],
such that the local bound is $\b[P\l]\leq0$, and the maximal
violation compatible with no-signaling is $\b[P\m]=1$.

As said above, local correlations can be generated with shared
randomness and local operations. In expression (\ref{local}), the
random variable $e$ stands for the information shared among the
parties, sometimes called {\em local hidden variable}. Depending
on its value, the $k\th$ party locally generates $P(a_k|x_k,e)$.
The distributions that cannot be written like (\ref{local}) are
called
nonlocal.

\subsection{Quantum correlations}

We call quantum those correlations that can be generated if the
parties share quantum information [entanglement], or equivalently,
those correlations that can be written as
\begin{equation}\label{quantum}
    P(a_1,\ldots a_n|x_1,\ldots x_n)
    =\mbox{tr}\!\left[ F_{a_1}^{(x_1)}\!\otimes\cdots\otimes
    F_{a_n}^{(x_n)} \rho \right],
\end{equation}
where $\rho$ is a quantum state, namely a unit-trace,
semi-definite positive matrix, and $\{F_{0}^{(x_k)},\ldots
F_{A_k-1}^{(x_k)}\}$ define what is called a {\em positive
operator valued measure} \cite{POVM}. That is, a set of positive
operators $\{F_{a_k}^{(x_k)}\}$ satisfying
$\sum_{a_k}F_{a_k}^{(x_k)}=\one,\,\forall\,x_k$.

\subsection{Isotropic correlations}

\begin{figure}
\begin{center}
  \includegraphics[width=10cm]{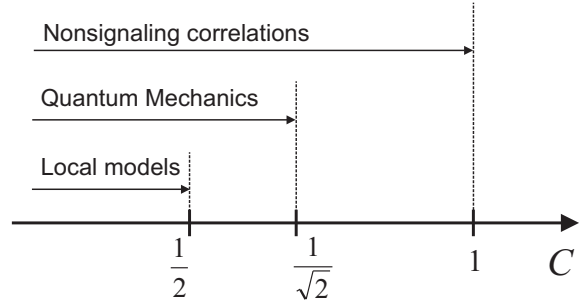}\\
  \caption{Value of $C$ for isotropic correlations
  (\ref{iso2}).}\label{chshfig}
\end{center}
\end{figure}

Let us define a particular family of bipartite distributions with
binary input/output. In the case where the marginal distributions
for $a$ and $b$ are unbiased, all the information of $P(a,b|x,y)$
is contained in the four correlation functions:
\begin{eqnarray}\label{}
    C_{xy}=&+& P(0,0|x,y)+P(1,1|x,y) \\
           &-& P(0,1|x,y)-P(1,0|x,y),\nonumber
\end{eqnarray}
for $xy=00,01,10,11$. One can always fix $C_{00},C_{01},C_{10}\geq
0$ by performing local reversible transformations. Once we have a
distribution in this canonical form, its nonlocality is decided by
the CHSH inequality \cite{chsh} in standard form:
\begin{equation}\label{chsh}
    \b_{\mbox{\tiny CHSH}}=\frac{1}{2}\left[
    C_{00}+C_{01}+C_{10}-C_{11} \right]-1.
\end{equation}
We call isotropic, denoted by $P\i(a,b|x,y)$, those correlations
with unbiased marginal distributions for $a$ and $b$ that satisfy
\begin{equation}\label{isoP}
    C_{00}=C_{01}=C_{10}=-C_{11} \geq 0.
\end{equation}
This family depends on a unique parameter $C=C_{00}$, whose
relation to the CHSH violation is
\begin{equation}
\b_{\mbox{\tiny CHSH}}[P\i]=2 C-1.
\end{equation}
In figure \ref{chshfig} we can see for which values of $C$ the
distribution $P\i$ belongs to the local and quantum set.
When $C=1$, this distribution is known as PR-box \cite{pr,NSC},
and is usually written as
\begin{equation}\label{iso}
    P_{\mbox{\tiny PR}}(a,b|x,y)=
\left\{
\begin{array}{ll}
    1/2 & \mbox{if }\ a+b \bmod 2=xy \\
    0 & \mbox{otherwise} \\
\end{array}%
\right. .
\end{equation}
This distribution can be considered the paradigm of nonlocal and
nonsignaling correlations (see \cite{conversion}). With this
definition, we can express any $P\i$ as the following mixture
\begin{equation}\label{iso2}
    P\i=C P_{\mbox{\tiny PR}}+
    (1-C) P^A_{\mbox{\tiny N}}P^B_{\mbox{\tiny N}},
\end{equation}
where $P^A_{\mbox{\tiny N}}$ is the local noise distribution for
Alice, independently of the inputs. Thus, one can interpret $C$ as
the probability of sharing a PR-box instead of local noise.

\section{Monogamy of nonlocal correlations}

While classical correlations can be shared among an indefinite
number of parties, it is well known that quantum correlations
cannot. This fact is often called {\em monogamy of entanglement}
\cite{monogamy}. In this section we prove that this is a generic
feature of all non-signaling theories.

First, let us recall a result already mentioned in \cite{NSC}.
{\em All Bell inequalities for which the maximal violation
consistent with no-signaling is attained by a unique distribution,
have monogamy constraints}. Suppose that $\b$ is a Bell inequality
with unique maximal violator $P\m$. If Alice-Bob maximally violate
this inequality $\b[P(a,b|x,y)]=1$, then, Alice and Clare are
completely uncorrelated. To prove this, first notice that because
all Bell inequalities $\b[P]$ are linear in $P$, $P\m$ must be an
extreme of the Alice and Bob polytope. Otherwise, the maximal
violator would not be unique. Second, using the definition of
marginal distribution and the no-signaling condition we have
\begin{eqnarray}
P\m(a,b|x,y) &=& \sum_c P(a,b,c|x,y,z)\nonumber\\
&=& \sum_c P(a,b|x,y,z,c)P(c|x,y,z)\nonumber\\
&=&\sum_c P(a,b|x,y,z,c)P(c|z),
\end{eqnarray}
for all $z$. But, because $P\m(a,b|x,y)$ is extremal, any such
decompositions must consist of only one term. This implies that
Clare is uncorrelated with Alice and Bob.

Actually, one can prove that all the CGLMP inequalities have a
unique nonsignaling probability distribution achieving its
algebraic maximum. This well-known set of inequalities was first
proposed in \cite{CGMLP} for the case of two inputs of $d$
possible outputs. One can easily see that imposing no-signaling
and maximal violation of CGLMP inequality identifies a unique
probability distribution $P(a,b|x,y)$. This means that this set of
Bell inequalities have the previous monogamy condition.

\subsection{$m$-shareability and locality}

Shareability represents a natural property in the analysis of the
monogamy of correlations. A bipartite probability distribution
$P(a,b|x,y)$ is said to be $m$-shareable with respect to Bob, if
there exists an $(m+1)$-partite distribution $P(a,b_1,\ldots
b_m|x,y_1,\ldots y_m)$ being symmetric with respect to
$(b_1,y_1)\cdots (b_m,y_m)$, with marginals $P(a,b_i|x,y_i)$ equal
to the original distribution $P(a,b|x,y)$. The following result
shows the relation between shareability and nonlocality.

\bigskip
{\bf Result 1:} {\em If $P(a,b|x,y)$ is $m$-shareable with respect
to Bob, then, it satisfies all Bell inequalities with $m$ (or less)
different values for the input $y$.}
\bigskip

{\em Proof:} To prove this statement, we construct a local model
for $P(a,b|x,y)$ when $y$ constrained to $y=1,\ldots m$ [without
loss of generality]. By assumption $P(a,b_1,\ldots
b_m|x,y_1,\ldots y_m)$ exists, then so $P(b_1,\ldots
b_m|y_1,\ldots y_m)$ and $P(a|x,b_1,\ldots b_m,y_1,\ldots y_m)$
do. In this local model, the information shared by the parties is
the string $(b_1,\ldots b_m)$, when the corresponding inputs are
fixed to $y_1=1, \ldots y_m=m$. Thus, using the definition of
conditional probabilities, we can decompose $P(a,b|x,y)$ in the
following way
\begin{eqnarray}\label{}
    P(a,b|x,y)= \sum_{b_1,\ldots b_m}
    P(b_1,\ldots b_m|1,\ldots m) \\
    \times P(a|x,b_1,\ldots b_m,1,\ldots m) \delta_{b,b_y}
    \nonumber.
\end{eqnarray}
Where the three factors in each term of the sum have to be
interpreted as the $P(e)$, $P(a|x,e)$ and $P(b|y,e)$ appearing in
the decomposition (\ref{local}), respectively.

Note that this result represents the extension of Theorem 2 in
\cite{terhal}, derived for quantum states, to the more general
nonlocal scenario. It also implies that if a state is
$X/Y$-shareable with respect to Alice/Bob, then it is local. In
particular, two-shareable states do not violate the CHSH nor the
CGMLP inequalities.

A converse of the previous result is also true: {\em if a state is
local, then it is $\infty$-shareable with respect to any party}.
To show the last statement, we explicitly construct the extension
[to $m$ Bobs] for the arbitrary local correlations written in
(\ref{local}):
\begin{eqnarray}\label{extension}
    P(a,b_1,\ldots b_m|x,y_1,\ldots y_m) \\
    =\sum_e P(e) P(a|x,e) P(b_1|y_1,e) \cdots P(b_m|y_m,e),
    \nonumber
\end{eqnarray}
with each distribution $P(b_i|y_i,e)$ being equal to the
$P(b|y,e)$ that appears in (\ref{local}). We can merge the
previous two statements in the following one:

\bigskip
{\bf Result 2: }{\em locality and $\infty$-shareability are
equivalent properties.}
\bigskip

This result is analogous to what happens in QM: a bipartite
quantum state is $\infty$-shareable if and only if it is separable
\cite{share}.

\subsection{Examples}

In what follows, we show that the CHSH inequality presents an even
stronger kind of monogamy.

\bigskip
{\bf Result 3: }{\em Consider a binary input/output tripartite
distribution $P(a,b,c|x,y,z)$. If Alice and Bob's marginal is
nonlocal, then Alice and Clare's marginal must be local.}
\begin{equation}\label{}
    \b_{\mbox{\tiny CHSH}}[P(a,b|x,y)]>0\ \Rightarrow\
    \b_{\mbox{\tiny CHSH}}[P(a,c|x,z)]\leq 0
\end{equation}
\bigskip

{\em Proof.} We prove this statement by contradiction. Suppose
that there exists a tripartite distribution $P(a,b,c|x,y,z)$ such
that both $P(a,b|x,y)$ and $P(a,c|x,z)$ are nonlocal. Then
Alice-Bob, and simultaneously Alice-Clare, can depolarize their
bipartite correlations and transform them into isotropic ones,
without decreasing the Bell violation. This procedure is shown in
Appendix B. Then, if Alice-Bob have larger $C$ than Alice-Clare,
Bob decreases it until both are equal (this procedure is explained
in Appendix A). An analogous thing is done in the opposite
situation. After this manipulations, both marginals are isotropic
and have the same value of $C$. This implies that the two
marginals are equal, and thus two-shareable. In section 3.3 we
have seen that, a two-shareable state cannot violate CHSH. This
finishes the construction of the contradiction. 

\bigskip

In more general situations strict monogamy no longer holds.
Indeed, one can easily design a situation where Alice shares a
PR-box with Bob, and another with Clare. This corresponds to a
case where Alice can choose between 4 inputs of 4 outputs, while
Bob and Clare are restricted to the simplest case of $Y=Z=B=C=2$.
Clearly, the corresponding Alice-Bob and Alice-Clare distribution
violate the CHSH inequality. A nicer and more symmetric example,
with only two inputs for each party, is given by the following
tripartite distribution
\begin{equation}\label{polygame}
    P^{ABC}=\frac{1}{2} P^{AB}_{\mbox{\tiny PR}\{0,1\}}
    P^C_{\mbox{\tiny N}\{0,1\}} +
    \frac{1}{2} P^{AC}_{\mbox{\tiny PR}\{2,3\}}
    P^B_{\mbox{\tiny N}\{2,3\}},
\end{equation}
where $P_{\mbox{\tiny PR}\{\alpha,\beta\}}$ is a PR-box with
outputs restricted to $a,b\in\{\alpha,\beta\}$, $P_{\mbox{\tiny
N}\{\alpha,\beta\}}$ is a local noise distribution with outputs
restricted to $a,b\in\{\alpha,\beta\}$, and the superindices label
the parties. In what follows, we prove that the Alice-Bob marginal
\begin{equation}\label{}
    P^{AB}=\frac{1}{2} P^{AB}_{\mbox{\tiny PR}\{0,1\}}+
    \frac{1}{2} P^{A}_{\mbox{\tiny N}\{2,3\}}
    P^B_{\mbox{\tiny N}\{2,3\}},
\end{equation}
is nonlocal. Assume the opposite: $P^{AB}$ can be expressed as a
mixture of local extreme points (\ref{local}). Because each local
extreme point has determined outcomes, we can split the local
mixture into a part with outcomes $\{2,3\}$, and a part with
outcomes $\{0,1\}$. The last, would correspond to a local
expansion of $P^{AB}_{\mbox{\tiny PR}\{0,1\}}$, but we know that
such thing does not exist. Now, using the symmetry of
(\ref{polygame}), we conclude that its marginals $P^{AB}$ and
$P^{AC}$ are both nonlocal.

\bigskip

In the case $X=Y=2$ and $A,B$ arbitrary, there is a situation
where strong monogamy still holds: where the reduced states of
Alice-Bob and Alice-Clare, consist both on isotropic correlations
with non-uniform noise [independent of the inputs]. First, let us
generalize the idea of isotropic distributions for arbitrary
output alphabets. The generalization of the PR-box is \cite{NSC}
\begin{equation}\label{prd}
    P_{\mbox{\tiny PR}}(a,b|x,y)=
\left\{
\begin{array}{ll}
    1/A & \mbox{if }\ a-b\bmod A=xy \\
    0 & \mbox{otherwise} \\
\end{array}%
\right. .
\end{equation}
In a natural way, we define
\begin{equation}\label{isod}
    P\i^{AB}=C P_{\mbox{\tiny PR}}^{AB}+
    (1-C) P^A_{\mbox{\tiny IND}} P^B_{\mbox{\tiny IND}},
\end{equation}
where $P^A_{\mbox{\tiny IND}}$ is an arbitrary local distribution
for Alice, independent of the inputs. It is clear that if Alice
and Bob add to their outputs a shared random number modulo $A$:
\begin{eqnarray}
    a \rightarrow a + r \bmod A \\
    b \rightarrow b+ r \bmod A ,
\end{eqnarray}
their distribution becomes:
\begin{equation}\label{}
    P\i^{AB} \rightarrow C P_{\mbox{\tiny PR}}^{AB}+
    (1-C) P^A_{\mbox{\tiny N}} P^B_{\mbox{\tiny N}},
\end{equation}
where $P^{A/B}_{\mbox{\tiny N}}$ is the (local) uniform
distribution independent of the inputs $x/y$. As in the case
$A=B=2$, if $C$ is positive, one of the parties can decrease its
value by performing a local operation. Using the same trick as
before, one can prove that all tripartite distributions where the
marginals Alice-Bob and Alice-Clare are both isotropic with
non-uniform noise (\ref{isod}), show strong monogamy.

\section{No-Cloning}

The Quantum No-Cloning Theorem represents one of the cornerstones
of Quantum Information Theory. It is usually explained as a
consequence of the nonorthogonality of quantum states and the
linearity of quantum time evolution. The relation between quantum
cloning and no-signaling has also been studied by several authors.
Indeed, if one assumes that (i) states are described by vectors in
Hilbert spaces, (ii) probabilities are obtained according to the
usual trace rule, and (iii) no-signaling, the optimal fidelity of
a cloning machine cannot be larger than the one allowed by quantum
dynamics \cite{gisin}. In what follows, we formulate the problem
independently of QM and show that

\bigskip

{\bf Result 4:} {\em All nonsignaling theories predicting the
violation Bell inequalities have a no-cloning theorem.}

\bigskip

A similar result was proved for the case of the CHSH inequality by
R. F. Werner, in \cite{qit}. Here we prove it for general nonlocal
theories, not necessarily violating the CHSH inequality. Suppose
that there exists a machine to which we can input a physical
system [in an arbitrary state], and it outputs two systems in
exactly the same state as the original one. We call such engine
{\em perfect cloning machine}. Let us consider the following
situation: Alice and Bob share the nonlocal distribution
$P(a,b|x,y)$, and perform the following two space-like separated
events. On one site, Alice chooses the input $x_0$ and obtains the
output $a_0$. On the other site, Bob performs $m$ clones of its
original system. For an observer who see first the event on
Alice's site, the description of Bob's input system is
$P(b|y,x_0,a_0)$. For this observer, Bob's system is completely
uncorrelated with the rest of the universe, and the functioning of
the perfect cloning machine is unambiguous:
\begin{equation}\label{cln}
    P(b|y,x_0,a_0) \rightarrow P(b_1,\ldots b_m|y_1,\ldots
    y_m,x_0,a_0)
\end{equation}
Obviously, the joint sate of all clones $P(b_1,\ldots
b_m|y_1,\ldots y_m,x_0,a_0)$ is such that when we trace all but
one, $P(b_i|y_i,x_0,a_0)$, this distribution is the same as the
original one, $P(b|y,x_0,a_0)$. Because we consider a perfect
cloning machine there is no distinction between pure and mixed
states: all are perfectly cloned. For an observer who first sees
Bob's operation, its description of the physical situation is
\begin{equation}\label{cln2}
    P(a,b_1,\ldots b_m|x,y_1,\ldots y_m)\ .
\end{equation}
But, because all descriptions must give consistent predictions,
the descriptions from the point of view of the two mentioned
observers (\ref{cln}) and (\ref{cln2}) must be the same, up to
conditioning on $a$. This implies that the original distribution
$P(a,b|x,y)$ is $m$-shareable. More concretely, because $m$ is
arbitrary, we can say that $P(a,b|x,y)$ is $\infty$-shareable.
According to the result of section 3.1, the original distribution
$P(a,b|x,y)$ must be local, in contradiction with the initial
assumption.

\subsection{Phase covariant cloning machine}

Once we have ruled out the existence of a perfect cloning machine,
it is interesting to look for the optimal imperfect one. Suppose
that its action is
\begin{equation}\label{}
    P(a,b|x,y) \quad\longrightarrow\quad P(a,b_1,b_2|x,y_1,y_2),
\end{equation}
where, without loss of generality we can assume that the final
distribution is symmetric with respect to $(b_1,y_1)$ and
$(b_2,y_2)$. By definition, the reduced distribution
$P(a,b_i|x,y_i)$ is two-shareable. This implies that it cannot
violate any two-input Bell inequality. In particular, if the
initial distribution $P(a,b|x,y)$ has $Y=2$, the resulting clones
are correlated with Alice's system in a local way.

Let us consider a particular case in the binary input/output
scenario. Consider that Alice and Bob share an isotropic
distribution with parameter $C$. Bob clones his subsystem, and,
according to the previous paragraph, the resulting clones are
locally correlated with Alice's subsystem. If we suppose that the
clones are isotropically correlated with Alice, the maximum value
for their parameter is $C_{\mbox{\tiny CLN}}=1/2$. Thus, the {\em
shrinking factor} associated to this cloning operation is
\begin{equation}\label{sf}
    \frac{C_{\mbox{\tiny CLN}}}{C}=\frac{1}{2\,C}.
\end{equation}
Now, consider the isotropic correlations that arise when measuring
a singlet with the observables that maximize the CHSH violation,
that is $P\i$ with $C=1/\sqrt{2}$. In this case, the shrinking
factor (\ref{sf}) coincides with the one of the {\em phase
covariant quantum cloning machine} $1/\sqrt{2}$
\cite{phasecovariant}, that is QM attains this maximum value for
the cloning of nonlocal correlations. In this sense, QM clones the
quantum correlations achieving the Cirelson bound in an optimal
way.

\section{Non-locality and privacy}

The monogamy of correlations and the impossibility of perfect
cloning seem immediately to be related to the concept of privacy.
If two honest parties know to share correlations with some degree
of monogamy, they can estimate and possibly bound their
correlations with a third dishonest party, the eavesdropper. In
this section we strengthen this intuitive idea, proving that under
the no-signaling assumption, a probability distribution contains
secrecy if and only if it is nonlocal. Recall that this does not
mean that this probability distribution can be transformed into a
secret key.

For the sake of simplicity we consider the bipartite case. In a
cryptographic scenario, one usually considers two honest parties
(Alice and Bob) each possessing a random variable $A$ and $B$, and
an eavesdropper (Eve) having $E$. The correlations among the three
random variables are described by a probability distribution
$P_{ABE}$ . On the other hand, it is meant by nonlocal
correlations those probability distributions conditioned on some
inputs $P(a,b|x,y)$ that cannot be written in the form of Eq.
(\ref{local}). It is in principle not so evident how to relate the
two scenarios. For instance, how to add (i) the third party in the
nonlocal scenario or (ii) the missing inputs for Alice and Bob in
the cryptographic scenario. Therefore, before proving the
equivalence between privacy and nonlocality one has to connect the
two considered scenarios.

\subsection{Secret correlations}

A tripartite probability distribution [without inputs] $P_{ABE}$
among two honest parties and an eavesdropper contains secrecy when
it cannot be generated by local operations and public
communication (LOPC), i.e. its formation requires the use of a
private channel or secret bits \cite{LOPC}. On the other hand,
$P_{ABE}$ can be generated by LOPC, if there exists a stochastic
map $E\rightarrow E'$ such that
\begin{equation}\label{LOPC}
    P_{AB|E'}=P_{A|E'}P_{B|E'}.
\end{equation}
We say that $P_{ABE}$ contains secrecy \cite{LOPC} when this is
not possible. We stress that this does not mean that many copies
of $P_{ABE}$ can later be used to obtain a secret key by LOPC.
Indeed, there are probability distributions with positive secrecy
content, which cannot be distilled into a secret key by LOPC
\cite{bound}.

Now, suppose Alice and Bob share a distribution $P(a,b|x,y)$. They
decide the inputs according to uniform distributions: $p(x)=1/X$
and $p(y)=1/Y$ \cite{noteinp}. Then, Alice's and Bob's information
is respectively $A=(a,x)$ and $B=(b,y)$. The random variables $A$
and $B$ are correlated according to
\begin{equation}\label{probab}
    P_{AB}=P(a,b|x,y)\frac{1}{XY} .
\end{equation}
Can Alice and Bob bound Eve's information on their outcomes from
their observed correlations? Can one prove that all possible
extension $P_{ABE}$ of $P_{AB}$, derived from $P(a,b|x,y)$ through
Eq. ({\ref{probab}), contain secrecy? This is of course impossible
if no assumption on the possible extensions are made. In general,
Alice and Bob can never exclude that Eve has a perfect copy of
their outcomes, unless some constraints are imposed. However, if
it is assumed that no faster-than-light communication is possible,
not all possible extension of the initial bipartite probability
distribution are allowed. Let us only consider extensions
$P(a,b,e|x,y)$ compatible with no-signaling. Thus, to each
$P(a,b|x,y)$ we can associate a family of tripartite distributions
\begin{equation}\label{ug}
    P_{ABE}=P(a,b,e|x,y)\frac{1}{XY},
\end{equation}
where $E=e$. We say that {\em $P(a,b|x,y)$ contains secrecy if all
its associated $P_{ABE}$ contain secrecy.}

\subsection{All nonlocal correlations contain secrecy}

The aim of this section is to show the link between the nonlocal
properties of $P(a,b|x,y)$ and the secrecy content of any possible
extension $P_{ABE}$, defined through (\ref{ug}). Before
proceeding, note that an equivalent way of defining local
correlations is as follows: a probability distribution
$P(a,b|x,y)$ is local (\ref{local}) when there exists a
[nonsignaling] extension $P(a,b,e|x,y)$ such that
\begin{equation}\label{local2}
    P(a,b|x,y,e)=P(a|x,e)P(b|y,e) .
\end{equation}
Now, assume one has a bipartite distribution $P(a,b|x,y)$ for
which there exists an extension $P_{ABE}$ with no secrecy content,
that is
\begin{equation}\label{L}
    P_{AB|E}=P_{A|E}P_{B|E}.
\end{equation}
Because processing the outcomes of a nonsignaling distribution gives
another nonsignaling distribution, any transformation $E\rightarrow
E'$ is included in the arbitrariness of the extension
$P(a,b,e|x,y)$. By using the definition of conditional
probabilities, one can see that (\ref{L}) is equivalent to
(\ref{local2}). That is, $P_{ABE}$ has no secrecy if and only if
there exists an extension of $P(a,b|x,y)$ satisfying (\ref{local2}),
which is to say that $P(a,b|x,y)$ is local. This establishes the
following equivalence.

\bigskip
{\bf Result 5: }{\em A distribution contains secrecy if and only if
it is nonlocal.}
\bigskip

It was already proven in \cite{crypto}, that all local
correlations (\ref{local}) can be distributed by LOPC. The public
message that one of the parties, say Alice, should send to the
rest in order to create the correlations, is precisely the
(hidden) variable $e$ that appears in (\ref{local}). Therefore, if
Alice and Bob's probability distribution is local, they cannot
exclude that the global probability distribution including Eve
does not contain any secrecy.

The following natural question is to identify those nonlocal
correlations distillable to a secret key and whether they can be
distributed using quantum states \cite{prep}. This will define
those quantum correlations secure against an eavesdropper only
limited by the no-signaling principle \cite{crypto}.

\section{Nonlocality and randomness}

We first start by showing that all nonlocal correlations have
random outcomes (see also \cite{pr}). Consider a deterministic
bipartite distribution $P\d(a,b|x,y)$. That is, $a$ and $b$ are
deterministic functions of $(x,y)$. Using this and no-signaling,
we can get the following equalities
\begin{eqnarray}
    P\d(a,b|x,y) &=& \delta_{(a,b),(f[x,y],g[x,y])}\nonumber\\
    &=& \delta_{a,f[x,y]}\ \delta_{b,g[x,y]}\nonumber\\
    &=& P(a|x,y)P(b|x,y)\nonumber\\
    &=& P(a|x)P(b|y).
\end{eqnarray}
The last line is a distribution of the form (\ref{local}).
Therefore, all deterministic distributions are local. Or in other
words, all nonlocal states have uncertain outcomes. This fact can
be straightforwardly extended to the $n$-party case. Thus, there
are two kinds of randomness in any nonsignaling theory with
nonlocal correlations. The first one reflects our ignorance and
corresponds to those probability distributions that can be written
as the convex combination of extreme points. But, like in QM,
there is also an intrinsic randomness even for extreme points, or
pure states. The PR-box (\ref{iso}) is an example of a pure state
with uncertain outcomes.

\subsection{Incompatible observables and uncertainty}

Finally, within QM it is said that two observables $(O_0,O_1)$ are
compatible if there exists a more complete one $O$ of which both
are functions: $(O_0,O_1)=f(O)$. Consider $P(a,b|x,y)$, we say
that the two observables in Bob's site $b_{0}$ and $b_{1}$
[corresponding to the inputs $y=0,1$] are compatible, if there
exists a joint distribution for both $P'(a,b_0,b_1|x)$. That is
\begin{eqnarray}
  \sum_{b_0} P'(a,b_0,b_1|x) &=& P(a,b_1|x,y=1)\ , \\
  \sum_{b_1} P'(a,b_0,b_1|x) &=& P(a,b_0|x,y=0)\ .
\end{eqnarray}
Or in other words, $P(a,b|x,y)$ is two-shareable with respect to
Bob if we restrict to $y=0,1$.

When the observables $(b_{0},b_{1})$ are not compatible, a
possible way of quantifying the degree of incompatibility is
\begin{eqnarray}\label{incomp}
    && \mbox{inc}[b_{0},b_{1}]= \min\!\big\{\eta>0: P(a,b|x,y)\\
    && =\eta P_{\mbox{\tiny INC}}(a,b|x,y)\,
    +(1-\eta)P_{\mbox{\tiny COM}}(a,b|x,y)\big\}, \nonumber
\end{eqnarray}
where $P_{\mbox{\tiny COM}}(a,b|x,y)$ is a distribution where
$b_{0}$ and $b_{1}$ are compatible, and, $P_{\mbox{\tiny
INC}}(a,b|x,y)$ is an arbitrary one. It is clear that the range of
$\mbox{inc}[b_{0},b_{1}]$ is $[0,1]$, and
$\mbox{inc}[b_{0},b_{1}]=0$ if and only if $b_{0}$ and $b_{1}$ are
compatible. In Appendix B it is proven that in the binary
input/output case, this minimization yields the CHSH violation:
\begin{equation}\label{}
    \mbox{inc}[b_0,b_1]=\b_{\mbox{\tiny CHSH}}[P(a,b|x,y)]\ .
\end{equation}

In the case of binary outputs or inputs, we are able to establish
a direct relation between $\mbox{inc}[b_{0},b_{1}]$ and the
uncertainty of $b_{0}$ and $b_{1}$:

\bigskip
{\bf Result 6: } {\em In the binary output case [$A=B=2$] the
following constraints hold:}
\begin{eqnarray}\label{heisenberg0}
    H(b_0)&\geq& h\!\left(\frac{1}{2}\mbox{inc}[b_0,b_1]\right),\\
    \label{heisenberg1}
    H(b_1)&\geq& h\!\left(\frac{1}{2}\mbox{inc}[b_0,b_1]\right),
\end{eqnarray}
{\em where $H(b)$ is the entropy of the output $b$, and $h(x)$ is
the binary entropy of $x$ \cite{entropy}. These inequalities also
hold in the binary input case [$X=Y=2$], and are still tight.}
\bigskip

The proof of this result is in Appendix B. Although this has the
flavor of the Heisenberg uncertainty relations, it differs in the
fact that here we do not have a trade off between the uncertainty
of each observable. In particular, if $b_0$ is deterministic,
inequality (\ref{heisenberg0}) implies $\mbox{inc}[b_0,b_1]=0$,
and hence, nothing prevents $b_1$ from being deterministic too. It
is also remarkable that, a deterministic observable is compatible
with any other.

Notice that in some of the proofs in this article, we express
distributions in terms of nonlocal extreme points. But, some
nonsignaling theories may not include them, like for example, QM
does not include PR correlations (\ref{iso}). It is important to
stress that this is not an inconvenient for the validity of the
proofs when applied to any particular theory. For instance,
although QM does not predict PR correlations we can always write
some quantum mechanical correlations as a mixture of PR and local
ones.

\section{Conclusions}

In this work, we have identified a series of features common to
all physical theories that do not allow for instantaneous
transmission of information, and predict the violation of Bell
inequalities. As shown, these two assumptions are sufficient to
prove:
    \begin{itemize}
    \item Constraints on how nonlocality is distributed among the
    correlations of different pairs of particles in multipartite scenarios.
    \item Impossibility of perfect cloning of states.
    \item Strict equivalence of the following properties:
        \begin{enumerate}
        \item nonlocality
        \item bounded shareability
        \item positive secrecy content
        \end{enumerate}
    \item A relation for the incompatibility of two observables and the uncertainty of their outcomes.
    \end{itemize}
Hence, some properties traditionally attributed to QM are generic
within this family of physical theories. For example: the fact
that two observables cannot be simultaneously measured on the same
system (incompatibility), becomes necessary to explain the
correlations observed in some experiments [violation of CHSH
\cite{aspect}], independently of the fact that we use models based
on noncommuting operators to explain such experiments (see also
\cite{qit}). Moreover, a no-cloning theorem can be derived without
invoking any nonorthogonality of states of linearity of the
evolution.

This indicates how constraining is the demand that a theory
compatible with special relativity predicts the violation of Bell
inequalities. One could actually say that there is not much room
left out of QM.

From a more fundamental point of view, this work proposes a
different approach to the study of quantum properties. In general,
QM has been studied in comparison with Classical Mechanics, that
is, starting from a more restrictive theory. Here, the idea is to
start from a more general family of theory, and to study
``quantum" properties common to all them. It is then an open
research project to identify those additional postulates that
allow one to recover the whole quantum structure.

\section{Acknowledgements}

We thank J. Barrett, S. Iblisdir, N. S. Jones, G.
Molina-Terriza, S. Popescu and V. Scarani for discussions. This
work is supported by the ESF, an MCYT ``Ram\'on y Cajal" grant,
the Generalitat de Catalunya, the Swiss NCCR ``Quantum Photonics",
OFES within the EU project RESQ (IST-2001-37559) and the U.K.
Engineering and Physical Sciences Research Council (IRC QIP).

\bigskip\bigskip

\section*{Appendix A. Depolarization and shrinking}

In this appendix it is shown that, in the case $X=Y=A=B=2$, any
distribution can be transformed into an isotropic one maintaining
the CHSH violation (\ref{chsh}) invariant. We call this process
{\em depolarization}. We also show that the parameter $C$ of an
isotropic distribution can be decreased with local operations. We
call this operation {\em shrinking}.

\bigskip
{\bf Depolarization:} this transformation can be implemented by
using 3 bits of shared randomness and local operations, in the
following two steps:

First step, Alice and Bob perform with probability $1/2$ one of
the following two operations:
\begin{enumerate}
    \item Nothing
    \item Flip $a$ and $b$
\end{enumerate}
This makes the correlations locally unbiased.

Second step, with probability $1/4$ both parties perform one of
the following four operations:
\begin{enumerate}
    \item Nothing
    \item Flip $a_{x=1}$ and $y$
    \item Flip $x$ and $b_{y=1}$
    \item Flip $x$, $a_{x=0}$, $y$ and $b_{b=1}$
\end{enumerate}
where flipping $a_{x=1}$ means that $a$ is only flipped when
$x=1$, that is $a \rightarrow a+x \bmod 2$. After the second step,
the resulting correlations satisfy (\ref{isoP}). It can be seen
that both steps keep invariant the violation of the CHSH
inequality.

\bigskip
{\bf Shrinking:} a useful observation is that when $C>0$,
the value of $\b_{\mbox{\tiny CHSH}}$ can always be decreased by
performing an operation in one site. This is accomplished when one
party, say Bob, outputs $b$ with probability $1-\epsilon$, and an
unbiased random bit with probability $\epsilon$. This operation
implements the transformation: $C\rightarrow (1-\epsilon)C$.

\section*{Appendix B. Proofs of section 6}

{\bf Result:} {\em In the case $A=B=X=Y=2$ the degree of
incompatibility of two observables is}
\begin{equation}\label{ruru}
    \mbox{inc}[b_0,b_1]=\b_{\mbox{\tiny CHSH}}[P]\ .
\end{equation}

{\em Proof.} The minimization in the definition of
$\mbox{inc}[b_0,b_1]$ in (\ref{incomp}), is completely equivalent
to the minimization of $p_{\mbox{\tiny NL}}$ in the optimal
eavesdropping extension (Appendix B). Then, we just have to
substitute $\mu$ by $p_{\mbox{\tiny NL}}$ which gives the equality
(\ref{ruru}).

\bigskip
{\bf Result 6: } {\em In the binary output case [$A=B=2$] the
following constraints hold:}
\begin{eqnarray}\label{h0}
    H(b_0)&\geq& h\!\left(\frac{1}{2}\mbox{inc}[b_0,b_1]\right),\\
    \label{h1}
    H(b_1)&\geq& h\!\left(\frac{1}{2}\mbox{inc}[b_0,b_1]\right),
\end{eqnarray}
{\em where $H(b)$ is the entropy of the output $b$, and $h(x)$ is
the binary entropy of $x$ \cite{entropy}. This inequalities also
hold in the binary input case [$X=Y=2$], and are still tight.}

{\em Proof.} Let us prove the above inequalities
(\ref{h0},\ref{h1}) for the binary output case. It is shown in
this case \cite{conversion} that, for all extreme points, the one
party marginals are deterministic or unbiased:
$\left[P(b=0|y),P(b=1|y)\right]\in\left\{[0,1],[1,0],[1/2,1/2]\right\}$.
In the next we see that, if one observable, say $y=0$, is
deterministic [$P(b_0|0)=0,1$] then it is compatible with all the
rest. To see this suppose that the outcome of $b_0$ is always
$b_0=\beta$, then, for any $y$, the joint distribution
$P(a,b_0,b_y|x,y)=P(a,b_y|x,y)\, \delta_{b_0,\beta}$ exists. Then,
$b_0$ and $b_y$ are compatible by definition. Now, let us
decompose $P_{\mbox{\tiny INC}}$ as a mixture of extreme points.
This mixture must not contain extreme points having the marginal
of $b_0$ or the marginal of $b_1$ deterministic. Otherwise, one
could move this extreme point to the mixture of compatible ones
$P_{\mbox{\tiny COM}}$, decreasing the value of $\eta$. Thus, the
marginals for $b_0$ and $b_1$ taken from $P_{\mbox{\tiny INC}}$
are always unbiased. Therefore, $\mbox{inc}[b_0,b_1]$ is the
probability of getting with certainty an unbiased outcome. The
situation where $b_0$ and $b_1$ have minimal entropy is when
$P_{\mbox{\tiny COM}}$ is deterministic. Suppose that
$P_{\mbox{\tiny COM}}(b=0|y=0)=1$, then recalling (\ref{incomp})
\begin{eqnarray}
    P(b=1|y=0) &=& \mbox{inc}[b_0,b_1]\,P_{\mbox{\tiny INC}}(b=1|y=0)
    \nonumber\\
    &=& \frac{1}{2}\mbox{inc}[b_0,b_1],
\end{eqnarray}
and thus the entropy of $b_0$ is
$H(b_0)=h(\mbox{inc}[b_0,b_1]/2)$. The same holds for $b_1$. In
general, when $P_{\mbox{\tiny COM}}$ is not deterministic, the
entropies will be larger than the bounds (\ref{h0},\ref{h1}).

Let us prove that the bounds (\ref{h0},\ref{h1}) also hold in the
case where inputs are binary, and the outputs belong to larger
alphabets. In that case, all extreme points have been classified
in \cite{NSC}. There, it is shown that, all extreme points have
local marginals where all outcomes with non-zero probability are
equiprobable. As discussed before, if we write $P_{\mbox{\tiny
INC}}$ as a mixture of extreme points, the marginals for $b_0$ and
$b_1$ given by these extreme points must have at least two
outcomes with nonzero probability. Otherwise the two observables
are compatible and we can attach the extreme point to
$P_{\mbox{\tiny INC}}$, decreasing $\eta$. The situation where
$b_0$ and $b_1$ have minimal entropy is when $P_{\mbox{\tiny
COM}}$ is deterministic, and $P_{\mbox{\tiny INC}}$ has only two
outcomes with nonzero probability for $b_0$ and $b_1$. In such
case, the inequalities (\ref{h0},\ref{h1}) are saturated. When
$P_{\mbox{\tiny INC}}$ has more than two outcomes with nonzero
probability for $b_0$ and $b_1$, the entropies will be larger.


\begin{thebibliography}{99}

\bibitem{Bell} J. S. Bell; Physics {\bf 1}, 195 (1964).

\bibitem{aspect} A. Aspect; Nature {\bf 398}, 189 (1999).

\bibitem{pr} S. Popescu and D. Rohrlich; Found. Phys. {\bf 24}, 379
(1994).

\bibitem{NSC} J. Barrett, N. Linden, S. Massar, S. Pironio, S.
Popescu and D. Roberts; Phys. Rev. A {\bf 71}, 022101 (2005).

\bibitem{infres}
W. van Dam, quant-ph/0501159; S. Wolf and J. Wullschleger;
quant-ph/0502030 \\
H. Buhrman, M. Christandl, F. Unger, S. Wehner and A. Winter;
quant-ph/0504133 \\
T. Short, N. Gisin and S. Popescu; quant-ph/0504134.

\bibitem{crypto} J. Barrett, L. Hardy and A. Kent; quant-ph/0404097.

\bibitem{chsh} J. F. Clauser, M. A. Horne, A. Shimony and R. A.
Holt, Phys. Rev. Lett. {\bf 23}, 880 (1969).

\bibitem{POVM} See for instance A. Peres, {\em Quantum Theory: Concepts and
Methods}, Kluwer, Dordrecht, (1995).

\bibitem{conversion} N. S. Jones and Ll. Masanes; quant-ph/0506182.

\bibitem{terhal} B. M. Terhal, A. C. Doherty and D. Schwab, Phys.
Rev. Lett. {\bf 90}, 157903 (2003).

\bibitem{gisin} N. Gisin; Phys. Lett. A {\bf 242}, 1 (1998).

\bibitem{qit} R. F. Werner; quant-ph/0101061.

\bibitem{entropy} The entropy of a probability distribution $P(a)$
is
\begin{equation}
H(a)=-\sum_a P(a)\log_2 P(a).
\end{equation}
The binary entropy function is
\begin{equation}
h(x)=-x\log_2 x -(1-x)\log_2 (1-x).
\end{equation}


\bibitem{share} R. F. Werner; Lett. Math. Phys. {\bf 17}, 359 (1989).

\bibitem{monogamy} V. Coffman, J. Kundu and W. K. Wootters, Phys.
Rev. A {\bf 61}, 052306 (2000); T. J. Osborne, quant-ph/0502176.

\bibitem{CGMLP} D. Collins, N. Gisin, N. Linden, S. Massar and S.
Popescu, Phys. Rev. Lett. {\bf 88}, 040404 (2002).

\bibitem{phasecovariant} C. -S. Niu and R. B. Griffiths, Phys. Rev. A
{\bf 60}, 2764 (1999); N. J. Cerf, J. Mod. Opt. {\bf 47}, 187
(2000).

\bibitem{LOPC} R. Renner and S. Wolf,
{\em Advances in Cryptology - EUROCRYPT 2003}, Lecture Notes in
Computer Science, Springer-Verlag, vol. 2656, pp. 562-577 (2003).

\bibitem{bound}
N. Gisin and S. Wolf, {\em Proceedings of CRYPTO 2000}, Lecture
Notes in Computer Science {\bf 1880}, 482, Springer-Verlag, 2000;
A. Acin, J. I. Cirac and Ll. Masanes, Phys. Rev. Lett. {\bf 92},
107903 (2004).


\bibitem{noteinp}
Actually all the results are independent of the choices $p(x)$ and
$p(y)$, if all the terms in these distributions are different from
zero.

\bibitem{AG}
A. Acin and N. Gisin, Phys. Rev. Lett. {\bf 94}, 020501 (2005).

\bibitem{prep}
Work in preparation.

\end{thebibliography}
\end{document}